\documentclass[fleqn]{ckm}                 

\usepackage{txfonts}            
\usepackage{pstricks}
\newcommand{\bea}{\begin{eqnarray}}
\newcommand{\eea}{\end{eqnarray}}

\newcommand{\bbar}{B_q^0 - \overline {B^0_q}}

\newcommand{\msbar}{{\overline{\rm MS}}}
\newcommand{\gev}{{\rm GeV}}
\newcommand{\mev}{{\rm MeV}}

\usepackage{epsf}

\confname{Workshop on the CKM Unitarity Triangle, IPPP Durham, April
  2003}

\title{Status of the computation of $f_{B_{s,d}}$, $\xi$ and $\hat g$}

\author{Damir Becirevic}
\address{Laboratoire de Physique Th\'eorique, Universit\'e Paris Sud, Centre
d'Orsay, F-91405 Orsay-Cedex, France}

\begin{document}

\begin{abstract}
Current status of the computation of the neutral $B$-meson mixing amplitudes, 
with particular attention to the heavy--light meson decay constants, is reviewed. 
The values for these quantities, as well as for the coupling of the pion to 
the lowest doublet of heavy--light mesons, are given. 
\end{abstract}

\maketitle

Decay constants of heavy--light mesons enter in the most direct way the 
standard analyses of the CKM unitarity triangle~\cite{hocker,yellowbook}, namely 
through the frequencies of mixing in the neutral $B$-meson systems:
\bea\label{UTA}
&&\Delta m_d = C_B m_{B_d} { f_{B_d}^2 \hat B_{B_d}} A^2 \lambda^6 
\left[ (1-\bar \rho)^2 + \bar \eta^2\right]\,,\cr
&&\hfill \cr
&&{\Delta m_d\over \Delta m_s} = {m_{B_d}\over m_{B_s}} 
\underbrace{ f_{B_d}^2 \hat B_{B_d}\over  f_{B_s}^2 \hat
B_{B_s}}_{\displaystyle{1/\xi^2}} \lambda^2 
\left[ (1-\bar \rho)^2 + \bar \eta^2\right]\,,
\eea
which, when combined with experimental values for the mass differences, 
constrain the vertex of the CKM  triangle, through the circle in the $\bar \rho-
\bar \eta$ plane. Pseudoscalar (and vector) meson decay constants, 
$f_{B_{d,s}}^{(\ast)}$, $f_{D_{u,s}}^{(\ast)}$, 
are also important in the studies of the corrections to the factorization 
approximation in non-leptonic $B$-decay modes. These constants are also important ingredients in 
the research of low energy physics effects coming from physics beyond Standard Model~\cite{nir}.
In what follows, I will focus on 
the computation of the pseudoscalar decay constants, although most of the discussion 
is equally applicable to the vector ones. I will briefly go through the novelties 
related to the bag-parameters $\hat B_{B_{d/s}}$, and close this short review by 
discussing the value of the phenomenologically important coupling of the pion to the 
lowest--lying doublet of heavy--light mesons.

\section{Strategies and difficulties to compute\\ the decay constants}

The decay constant $f_H$ of the heavy--light pseudoscalar meson $H$ is defined through
\bea\label{def:fB}
&&\langle 0\vert \bar Q \gamma_\mu \gamma_5 q \vert H(p) \rangle =  i p_\mu f_H\,,
\eea
where $Q$ is either $c$- or $b$-quark, and 
$q$ is $s$- or $u/d$-quark.~\footnote{
The mass difference between $m_d-m_u$ quarks cannot be resolved by any of the methods 
to compute the non-perturbative effects.} 
Although the matrix element~(\ref{def:fB}) is the simplest one, the fact that 
one quark is heavy and the other is light makes its computation extremely difficult 
(in spite of the simplifications stemming from the heavy quark and chiral symmetries).

In various quark models the decay constant is either a parameter of the model, or 
its value depends crucially on the specific choice of 
the model's parameters. This situation is clearly unacceptable as we seek 
a ``model-independent" estimate (an {\it ab initio} determination), solely relying  
on the underlying theory, on QCD. A step in this direction was made by employing 
the duality sum rules, both in QCD and in heavy quark effective theory 
(HQET).~\footnote{See ref.~\cite{jamin} for the most 
recent update and for the full list of references.} 
The non-perturbative contributions in this approach are parametrized by the power 
corrections, the coefficients of which are the vacuum condensates. Apart from the chiral condensate, 
the essential non-perturbative input in evaluating $f_B$ comes from the gluon and the mixed 
quark-gluon condensates. Those two are either not well defined or their values are poorly known. 
For example, the gluon condensate, as estimated from the comparison of the sum rules with the corresponding 
experimental information on the $\rho^{(n)}$ resonances~\cite{geshkenbein} 
and charmonium resonances~\cite{zyablyuk}, differ by a factor of roughly $8$: 
\bea
\left< {\alpha_s\over \pi} G_{\mu\nu}^2\right> = \biggl.(0.074\pm 0.023)\ \gev^4\biggr|_{``\rho"}\!\!,
\biggl.(0.009\pm 0.007)\ \gev^4\biggr|_{``\psi"}\!\!.\nonumber
\eea
Besides, it is often unclear where the onset of the quark--hadron duality in the sum rule takes place (parametrized 
by the threshold parameter $s_0$). In short, the benefit of the method is that it made a step towards the first
principle QCD calculations of the hadronic quantities; the drawback is the existence of too many parameters 
whose values are loosely constrained, thus prohibiting the precision determination of the hadronic quantities, and
of $f_B$ in particular.

Lattice QCD is the closest to our ultimate goal, first-principle QCD calculation of the matrix 
element~(\ref{def:fB}). Although the calculations are based on the numerical simulations of the QCD vacuum fluctuations, 
the great bonus is that the only parameters appearing in the computations are those that appear in the QCD Lagrangian, 
namely the bare strong coupling and  quark masses. A strategy to compute $f_B$ on a 4-dimensional lattice ($L^3\times T$) 
 is very simple and can be summarized in $4$ steps:
\begin{itemize}
\item[1.] Generate an SU(3) gauge field configuration ${\cal U}$ (by using the Monte Carlo technique);
\item[2.] For each time slice, $t\in [0,T)$, in the background field produced in 1, 
calculate the correlation function,
\bea \label{correlator}
&&\sum_{\vec x}\langle 0\vert \overbrace{\bar Q(x) \gamma_0 \gamma_5 q(x)}^{\displaystyle{A_0(x)}}A_0^\dagger(0)\vert 0\rangle .
\eea 
\item[3.] Repeat steps 1 and 2, a number $N_{\rm conf.}$ of times, $N_{\rm conf.}$ refering to the number 
of independent gauge field configurations. From the average over the statistical sample, extract $f_H$ as 
\bea
\langle \sum_{\vec x} A_0(x)
A_0^\dagger(0)\rangle_{\cal U}& \stackrel{t\gg 0}{=}&
{\vert \langle 0 \vert \bar Q \gamma_{_0}
\gamma_{_5} q \vert H\rangle \vert^2  \over 2 m_H} e^{- m_{H} t}\, +\,
\dots \nonumber\\
&=& {1\over 2} { f_H^2} m_H e^{- m_{H} t}\, +\,
\dots\;,
\eea
where the (euclidean) time is large enough for the higher (heavier) excited states 
to be indeed suppressed w.r.t. the lowest lying one; 
\item[4.] Repeat 2 and 3 for several different light and heavy quark masses, $m_q$ and $m_Q$ respectively.
\end{itemize}
This simple procedure is very demanding in practice: to avoid large lattice artefacts, while  
working with reasonably light quarks, the size of the lattice box ($L$) must be very large. Simultaneously,  
and to resolve the propagation of the heavy quark, a very small lattice spacing (``$a$") is needed.
Compared with the physical quark masses, the currently available computers allow us to work with
\bea
&&m_c \leq m_Q < m_b\,,\ {\rm and} \quad m_{u/d} < m_q \leq m_s \;.
\eea 
In other words, we can directly compute only the $f_{D_s}$ decay constant.

\begin{figure}[tbh]
\vspace*{-2mm}\centerline{\hspace*{-.5mm}\epsfxsize=0.54\textwidth\epsffile{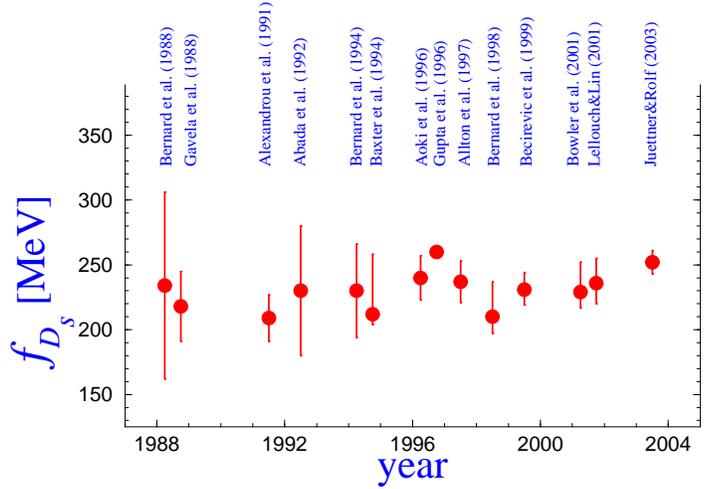}}
\caption[]{Evolution of the lattice estimates of $f_{D_s}$, as obtained by various lattice groups. 
List of papers corresponding to the above results, organized chronologically, can be found in 
refs.~\cite{fDsref,LL}. The last result, from ref.~\cite{rolf1},  
is the final lattice value obtained in the quenched approximation. }
\label{fig1}\end{figure}
In fig.~\ref{fig1} we present the evolution of the results obtained from the lattice QCD 
simulations over the past 15 years. All results are obtained in the quenched
approximation. The last value is actually the final quenched lattice estimate for this quantity. It is  
obtained after implementing all the important elements of theoretical progress 
made in the 90's, namely: 
(i) the quark action and the axial current are improved, 
providing a better scaling to the continuum limit; 
(ii) the local axial current defined on the lattice is matched to 
its continuum counterpart non-perturbatively; (iii) the simulations 
are made at several (four) small lattice spacings ``$a$", which allows for 
a smooth extrapolation to the continuum limit ($a\to 0$). The final result of ref.~\cite{rolf1} is
\bea
&&f_{D_s} = 252\pm 9\ \mev.
\eea

\begin{figure}[tbh]
\vspace*{-3mm}\centerline{\hspace*{-5.5mm}\epsfxsize=0.5\textwidth\epsffile{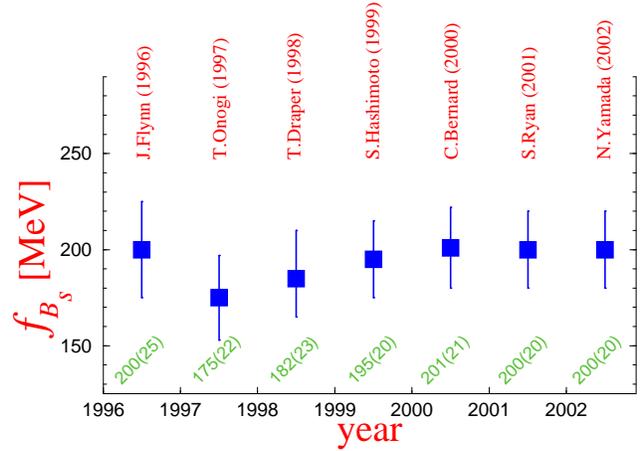}}
\caption[]{Evolution of the world average of the (quenched) lattice estimates of $f_{B_s}$, reported at the 
annual  ``Lattice" meetings~\cite{raporteurs}.  In the average enter the results obtained by all the strategies to treat the heavy 
quark on the lattice that are discussed in the text.}
\label{fig2}\end{figure}

The fact that the $b$-quark cannot be resolved on the available lattices opened three options: 
\begin{itemize}
\item[(a)] Compute $f_H$ for the accessible $m_Q$ and then from the fit of $\Phi(m_Q) = C(m_Q) f_H \sqrt{m_H}$,  
in $1/m_H$, extrapolate to the $B$-meson mass. $\Phi(m_Q)$ is the quantity that scales with the inverse 
quark(meson) mass as a constant. The sizable $1/m_H$ corrections are determined from the lattice data. 
The resulting values, however, have large errors, mainly due to uncertainty of whether or not one includes 
the terms of ${\cal O}(1/m_H^{n\geq 2})$ in the extrapolation. Besides, the larger quark masses may induce 
large lattice cut-off artefacts that are hard to quantify. 
\item[(b)] Difficulties in controlling the systematic errors of the strategy (a) incited many groups to 
attempt treating the heavy quark in an effective theory. 
Computation of the matrix element~(\ref{def:fB}) in the static limit of HQET~\cite{eichten} turned out 
to be very difficult, mainly because of the very poor signal~\cite{hashimoto}. That problem (of the poor signal) was 
circumvented by including the ${\cal O}(1/m_Q)$ corrections in both the Lagrangian and in the operator (axial current) by 
discretizing the NRQCD~\cite{lepage}. In spite of its benefits (working with large quark masses) this method cannot  
be used for the precision computation since, on the lattice, the $1/m_Q$ terms become ${\cal O}[1/(am_Q)]$, 
where $``a"$ is the small lattice spacing. In other words, the continuum limit does not exist. Besides, 
in both approaches (HQET and NRQCD) the non-perturbative renormalization of the axial current was not 
feasible. 
The third effective approach has been developed by the Fermilab group~\cite{EKM}. It consists of pushing 
 the propagating heavy quark, $Q$, over the lattice cut-off and then expanding in powers of $1/m_Q$ [not $1/(am_Q)$]. 
The key in that procedure is to match the relativistic with 
non-relativistic energy-momentum dispersion relations, where the mismatch 
is accounted in distinguishing the masses that appear in the non-relativistic expansion as $M_{``static"}$, 
$M_{``kinetic"}$, and so on. That matching is typically made non-perturbatively 
while the renormalization is made only perturbatively.~\footnote{For details and 
a complete list of references, please see refs.~\cite{kronfeld-handbook}. }
\item[(c)] Combine the value of $\Phi(\mu)$ obtained in the static limit ($1/m_Q=0$), with the decay constants 
computed with the directly accessible heavy--light mesons, and {\it interpolate} to $1/m_b$. 
The main obstacles in following this strategy, as mentioned above, are the poor statistical quality of the static 
correlation functions and the missing non-perturbative evaluation of the renormalization constant.
\end{itemize}
The overall agreement among results for $f_{B_s}$, as obtained by using various strategies, was quite impressive. 
However, none of the approaches was fully satisfactory to provide a precision measurement 
(even within the quenched approximation). This is why the errors remained essentially unchanged over 
the last 7 years, as can be seen in fig.~\ref{fig2}, where we plot the evolution of the world 
average of the quenched lattice estimates of $f_{B_s}$~\cite{raporteurs}.

\section{New ways to get $f_{B_s}$ (quenched)}

The two results, reported this year, actually supersede all the previous quenched calculations.

\subsection{Method of finite volumes}

To simulate the $b$-quark directly on the lattice, one can actually set a very small lattice spacing, but 
then the physical volume becomes too small to accommodate the propagation of the light quark. 
To get around this problem, the ``Tor Vergata" group~\cite{tov} proceed as follows: (i) in a 
small box of side $L=L_0$, they compute $f_H(L_0)$; (ii) they double 
the side of the box, at fixed lattice 
spacing $a$, recompute  $f_H(2 L_0)$, and take the ratio, $\Sigma_2(a) = f_H(2 L_0)/f_H(L_0)$; (iii) step (ii) is 
then repeated $n$ times until $\Sigma_{2 n}(a)\to 1$; (iv) the whole procedure is redone for several 
values of the lattice spacing $a$, allowing the smooth continuum extrapolation, $\Sigma_{2 n}(a)\to \Sigma_{2 n}(0)\equiv\sigma_{2 n}$.
Throughout the calculation, they set $T = 2 L$, and extract the decay constant 
precisely at $T/2$.

The actual calculation with this clever procedure was recently completed in ref.~\cite{nazario}. 
Let us first focus on $f_{D_s}$. In the first step, they work with $L_0 = 0.4$~fm, so that 
at $t=L_0$ they certainly cannot extract $f_{D_s}(L_0)$, but rather 
\bea
f_{D_s}^{\rm eff.}(L_0)=f_{D_s} (L_0) \left[ 1 + \sum_n 
\sqrt{m_{D_s^{(n)}}f_{D_s^{(n)}}^2\over m_{D_s} f_{D_s}^2} e^{-(m_{D_s^{(n)}}-m_{D_s}) L_0}  \right] \,,
\eea
where the sum runs over all excitations that are not heavy enough to be exponentially suppressed at 
 (small) $L_0$. After doubling the size, $L_0\to 2 L_0$, the exponential suppression will be more efficient, 
so that after $2$ or $3$ steps all excitations will die out. After performing the continuum extrapolation in each step, 
they obtain 
\bea
f_{D_s}^{\rm eff.}(L_0)= 644 \pm 3\ \mev,&& \sigma_2 ={f_{D_s}^{\rm eff.}(2L_0)\over f_{D_s}^{\rm eff.}(L_0)}=0.414(3),\cr
&&\sigma_4 ={f_{D_s}^{\rm eff.}(4 L_0)\over f_{D_s}^{\rm eff.}(2 L_0)} = 0.90(2),\cr
&&\hspace*{-32mm} \Longrightarrow \ f_{D_s}=f_{D_s}^{\rm eff.}(L_0) \sigma_2 \sigma_4 = 240(5)(5)\ \mev\,,
\eea
thus in good agreement with the value obtained in ref.~\cite{rolf1}. 
The main reason why $\sigma_{2}$ is so different from $1$ seems to be the presence of the 
non-decoupled excited states. This can be seen from what the authors of ref.~\cite{nazario} present: they show that 
$\Sigma_{2 n}(a)$ depends only weakly on both the heavy and light quark masses, and they verify only a tiny 
dependence on the lattice spacing.

When working with the $B_s$-meson, $f_{B_s}^{\rm eff.}(L_0)$ can be computed directly, but doubling 
the side of the box to compute $\sigma_{2n}$ becomes unfeasible. Instead, one can afford to compute 
$\sigma_{2n}(m_Q)$, for $m_c \lesssim m_Q \lesssim m_b/2$, and then reach $\sigma_{2n}(m_b)$, through 
an extrapolation. For that purpose they assume the heavy quark expansion in the finite box, so that 
\bea\label{ansatz}
&&\sigma_{2n}(m_Q)= \alpha_{2n} + \beta_{2n}/m_Q\,,
\eea
where the constant $\beta$ naively scales as $1/(2^{n-1}L_0)$. Their data indeed verify eq.~(\ref{ansatz}), leading to
\bea
f_{B_s}^{\rm eff.}(L_0)= 475 \pm 2\ \mev,&& \sigma_2(m_b) =0.417(3),\cr
&&\sigma_4(m_b) = 0.97(3),\cr
&&\hspace*{-44mm} \Longrightarrow \  f_{B_s}=f_{B_s}^{\rm eff.}(L_0) \sigma_2(m_b) \sigma_4(m_b) = 192(6)(4)\ \mev\,.
\eea
The second error indicates the combined systematic uncertainty due to renormalization constants and to the 
continuum extrapolations.

\subsection{Combining with the static $f_{B_s}$}

As we mentioned before, the value of the decay constant in the static limit was plagued by two main 
difficulties: 1. the renormalization of the axial current in HQET on the lattice; 2. bad signal-to-noise ratio.
Both problems have been solved recently.

By a judicious choice of the renormalization condition, 
the authors  of ref.~\cite{sommer}  provided 
a solution to the first problem (see around eq.(2.15) of that paper). They devised the method 
of subtracting the power-divergent residual mass counterterm non-perturbatively, which then allowed them to 
apply the standard non-perturbative renormalization procedure in the Schr\"odinger functional scheme~\cite{luscher}.

The solution to the second problem came recently too~\cite{michele}. In the static limit, 
eq.~(\ref{correlator}) becomes
\bea
&&C^{\rm cont.}(t) = \langle {\rm Tr}\left( {1+\gamma_0\over 2} {\cal S}_q(\vec 0,t) 
{\cal P}{\rm e}^{ig \int_0^t  d \tau A_0(\vec
0,\tau)}
\right) \rangle \cr
&&\rightarrow C^{\rm latt.}(t) = \langle {\rm Tr}\left( {1+\gamma_0\over 2} {\cal S}_q(\vec 0,t) 
\prod_{\tau=1}^{t-1} U_0(\tau)
\right) \rangle\ ,
\eea
where ${\cal S}_q(\vec 0,t)$ is the light quark propagator at $\vec x=0$, 
while the Wilson line on the lattice is simply the time-ordered  product of 
 link variables. Empirically, the statistical qua\-li\-ty 
of $C^{\rm latt.}(t)$ becomes much better if some kind of ``fattening" of 
the link variable is made: $U_0(\tau) \to U_0^{\rm fat}(\tau)$.~\footnote{
Fat link is obtained by averaging over the ``staples" made of links that are the 
first neighbours to a link which is being fattened. Among various recipes to do that, 
the most efficient appears to be the one proposed in ref.~\cite{knechtli}.}

After extrapolating to the continuum limit and converting the static HQET result to QCD, 
from ref.~\cite{michele} we learn that, 
\bea
&& f_{B_s}^{\rm stat} = 225 \pm 16 \ \mev .
\eea
That result has then been combined with those that are accessible directly, 
with the propagating heavy quark of mass $(a m_Q) \lesssim 0.65$ (in lattice units) as to avoid larger 
discretization errors. After extrapolating to the continuum at each of the simulated heavy quarks, 
they show a very smooth linear behaviour of the combined set of data. Their preliminary result, 
presented in ref.~\cite{rolf2}, is:
\bea
&&f_{B_s} = 206 \pm 10 \ \mev .
\eea
The bonus of that result is obviously a much better precision, since the extrapolation to $b$-quark is now 
replaced by the interpolation. Notice however that the slope (i.e. $1/m_Q$-dependence) of $\Phi_Q\equiv 
C f_{H_s}\sqrt{m_{H_s}}$, quoted in ref.~\cite{rolf2}, is completely compatible with the previous 
calculations in which the strategy with  propagating heavy quark was used at a fixed value of the lattice 
spacing.

\subsection{Status and perspectives of quenched $f_{D_s}$ and $f_{B_s}$}

After $15$ years of computing the heavy--light decay constants on the lattice, 
in the quenched approximation, we are finally in a position to quote 
the precise values. From the results discussed in this section, I conclude
\bea\label{quenched}
&&f_{D_s}^{N_{\rm f}=0}=245\pm 6\ \mev\,,\quad 
f_{B_s}^{N_{\rm f}=0}=194\pm 6\ \mev\,.
\eea
It should be stressed that these values are obtained by fixing the lattice 
spacing to $r_0=0.5$~fm, which, however, is an assumption. Using other quantities, 
such as $f_K$, $m_{K^\ast}$, leads to different values, which amount to 
a systematic error of about $5\%
$. The above results are obtained by working with fine-grained lattices, 
by implementing the non-perturbative renormalization and after a smooth extrapolation  
to the continuum limit. Alpha group plans to attempt a computation of the $1/m_b$ 
corrections to the $f_{B_s}^{\rm stat}$, to check whether these are enough to feed the linear 
dependence that they observed after having combined the static and the results 
obtained with relativistic (propagating) heavy quarks~\cite{rolf2}. 

The most challenging issue that remains to be studied and understood is to assess 
the error induced by the quenched approximation. The knowledge gained about that systematics 
is what we discuss next.

\section{Unquenching $f_{D_s}$ and $f_{B_s}$}

To unquench the lattice results means that one has to include the effects of dynamical 
quarks (those popping up through loops in the background gauge field).
To incorporate the light dynamical quarks is, in principle, possible, but very costly in practice.  
The most tested and widely used robust algorithm, Hybrid Monte Carlo (HMC)~\cite{HMC}, allows one to include 
two light degenerate quarks of the mass close to the strange quark (please see~\cite{HMC2}). 
Getting to the quarks lighter than half of the strange quark mass is 
nowadays impossible, if we use the standard  Wilson quark action and keep the finite volume 
effects under control. To get over that limit, as well as 
to include the third (strange) dynamical quark, a substantial 
progress in algorithm building is badly needed. An important step in that 
direction has been made in ref.~\cite{montvay}.

Up to now, the partially (un)quenched computations are 
made by using the Wilson quark action for the light and one of the effective approaches 
to treat the heavy quark. 

By confronting the results of quenched and unquenched studies (with $N_{\rm f}=2$ degenerate sea quarks)  
obtained by using the NRQCD treatment of the heavy quark on the lattices with 
$a^{-1} \approx 2$~GeV~\cite{CPPACS1,JLQCD1,JLQCD2}, the following effect of the dynamical quarks 
can be deduced :
\bea
{\rm CP-PACS}\!\!\!\!\!\!&:&{f_{B_s}^{N_{\rm f}=2}\over f_{B_s}^{N_{\rm f}=0}} = 1.10(5)\,,\cr
{\rm JLQCD}\!\!\!\!\!\!&:&{f_{B_s}^{N_{\rm f}=2}\over f_{B_s}^{N_{\rm f}=0}} \simeq 1.13(5)\,.
\eea
With the Fermilab treatment of the heavy quark, instead, one has~\cite{CPPACS2,milc}:~\footnote{The MILC group made the 
 most extensive unquenched ($N_{\rm f}=2$) study of the heavy--light decay constants. 
 Please see ref.~\cite{milc} for the detailed report on their findings.}
\bea
{\rm CP-PACS}\!\!\!\!\!\!&:&{f_{B_s}^{N_{\rm f}=2}\over f_{B_s}^{N_{\rm f}=0}} = 1.14(5)\,,\quad 
               {f_{D_s}^{N_{\rm f}=2}\over f_{D_s}^{N_{\rm f}=0}} = 1.07(5)\,,\cr
{\rm MILC}\!\!\!\!\!\!&:&{f_{B_s}^{N_{\rm f}=2}\over f_{B_s}^{N_{\rm f}=0}} \simeq 1.09(5)\,,\quad 
            {f_{D_s}^{N_{\rm f}=2}\over f_{D_s}^{N_{\rm f}=0}}\simeq 1.08(5)\,.
\eea
The average of the above results, combined with the values given in eq.~(\ref{quenched}), 
leads to
\bea\label{nf=2}
&&{f_{B_s}^{N_{\rm f}=2}\over f_{B_s}^{N_{\rm f}=0}} = 1.12(5) \,\Rightarrow \,
 f_{B_s}^{N_{\rm f}=2}=217\pm 12\ \mev,\cr
&&{f_{D_s}^{N_{\rm f}=2}\over f_{D_s}^{N_{\rm f}=0}} = 1.08(5) \,\Rightarrow \,
 f_{D_s}^{N_{\rm f}=2}=265\pm 14\ \mev.
\eea
Unaccounted for is $\sim 5\%
$ of systematic uncertainty, which we should keep in mind, due to the scale setting when converting from the lattice to the physical units. 
A detailed study of that uncertainty in the unquenched ($N_{\rm f}=2$) simulations has not been 
made so far.

Finally, we are interested in the situation in which $N_{\rm f}=2+1$, where the extra flavour 
would correspond to the strange sea quark. With Wilson fermions such a study is extremely expensive, 
although the algorithms to do such simulations already exist~\cite{algorithms}. Instead, 
a study with the so-called staggered light quarks has been made~\cite{wingate}. Although the 
method is relatively cheap to implement, one of the unsatisfactory features can be formulated as follows. 
With the staggered action,  
each dynamical quark flavour on the lattice comes in four ``tastes" (copies), and it is unclear 
how one can relate such Dirac determinant $\Delta_{\rm staggered}$, to the desired  
one. A current practice of taking 
$+\left(\Delta_{\rm staggered}\right)^{1/4}$ induces non-localities that are potentially 
problematic, and the whole formulation may not correspond to QCD. 
The proponents of the method will hopefully 
study this issue more carefully. Having that off my chest, I can now compare the quenched and unquenched 
results of ref.~\cite{wingate}, in which the NRQCD treatment of the heavy quark has been used, to quote 
\bea
&&\left(f_{B_s}^{N_{\rm f}=2+1}/ f_{B_s}^{N_{\rm f}=0}\right) \approx 1.15\,,
\eea
i.e. fully consistent with the result in eq.~(\ref{nf=2}).

In conclusion, the current values of the pseudoscalar decay constants, with the light quark $s$, are:

\psshadowbox[fillcolor=yellow,linewidth=.0cm]{{\parbox[c]{8.1cm}{
\bea
&&f_{B_s} = 217\pm 12\pm 11\ \mev,\cr
&& \hfill \cr
&&f_{D_s} =  265\pm 14\pm 13\ \mev.
\eea}}}

\section{$f_{B_s}/f_{B_{u/d}}$ and $f_{D_s}/f_{D_{u/d}}$}

So far we discussed the case with the strange light quark in the heavy--light meson. In particle physics 
phe\-no\-me\-no\-lo\-gy, however, more needed is the information  about $f_{B_{u/d}}$, $f_{D_{u/d}}$, and about 
the SU(3) breaking ratios, $f_{B_s}/f_{B_{u/d}}$ and $f_{D_s}/f_{D_{u/d}}$. The best known example of how important is 
the knowledge of these quantities is the one given in eq.~(\ref{UTA}).

As we already pointed out in the previous sections, the physical $m_{u/d}$ is too small 
to be simulated directly on the lattice. Currently feasible are the light quarks whose mass,  
with respect to the physical strange quark mass ($r = m_q/m_s^{\rm phys.}$), lie  between
\bea\label{rlatt}
&&1/2 \ \lesssim \ r \ \lesssim 2\,,
\eea
implying the necessity for quite a long extrapolation  to 
the physically relevant limit, $r\to r_{u/d}=0.04$~\cite{leutwyler}. 
This is where the staggered quark action has 
a great advantage over the standard (Wilson) one: due to the explicit chirality, 
with staggered quarks one can reach quarks as light as $r \approx 1/4$~\cite{wingate}, 
implying a  better control over the extrapolation to $r_{u/d}$.

Within the accessible range of the light quark masses~(\ref{rlatt}), the results of 
both quenched and unquenched simulations (with $N_{\rm f}=2$) suggest a pronounced 
linear dependence on the light quark mass,  
\bea\label{latt-slope}
&&{f_{B_s}\over f_{B_q}}\ =\ 1\ +\ X_b \cdot (1-r)\,,
\eea 
with little or no room for a term $\propto r^2$. 
For the quenched value of the slope $X_b$, I will take the 
one quoted in ref.~\cite{nazario}, namely 
$X_b^{N_{\rm f}=0} = 0.13(2)(1)$. To get the unquenched slope, 
we first compile the available results for the SU(3) breaking ratio 
as computed with $N_{\rm f}=2$, divided by its quenched value, where 
both  results are being obtained from an extrapolation of 
the form~(\ref{latt-slope}). 
\begin{figure}[tbh]
\vspace*{-5mm}\centerline{\epsfxsize=0.4\textwidth\epsffile{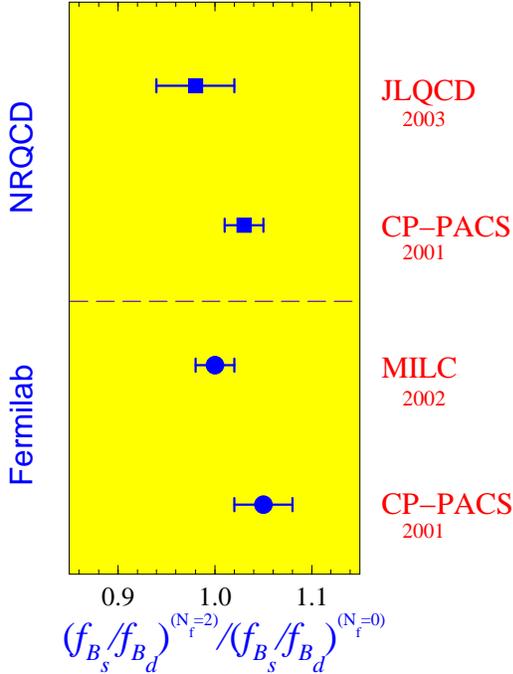}}
\caption[]{Ratio of the slopes which appear in eq.~(\ref{latt-slope}) in light quark mass, 
as obtained from the quenched and unquenched simulations, by using the NRQCD and the Fermilab  
treatment of the $b$-quark. From up to down, the results correspond to refs.~\cite{JLQCD1,JLQCD2},\cite{CPPACS1}, 
\cite{milc} and \cite{CPPACS2}.}
\label{fig3}\end{figure}
Those results are plotted in fig.~\ref{fig3}. The average is~\bea
&&{\; (f_{B_s}/f_{B_d})^{N_{\rm f}=2}\over (f_{B_s}/f_{B_d})^{N_{\rm f}=0}} = 1.02\pm 0.03\,,
\eea
or, the SU(3) breaking ratio is only slightly sensitive 
to the switch from $N_{\rm f}=0 \to 2$, for the 
light quark mass in the range indicated in eq.~(\ref{rlatt}).  
Notice also that a preliminary study with staggered fermions indicates 
that this feature persists even after going to ${N_{\rm f}=2+1}$.
From the above discussion we can now easily extract the slope: 
\bea
&&X_b^{N_{\rm f}=2} = 0.15(4)\,.
\eea

\begin{figure}[tbh]
\vspace*{-5mm}\centerline{\epsfxsize=0.5\textwidth\epsffile{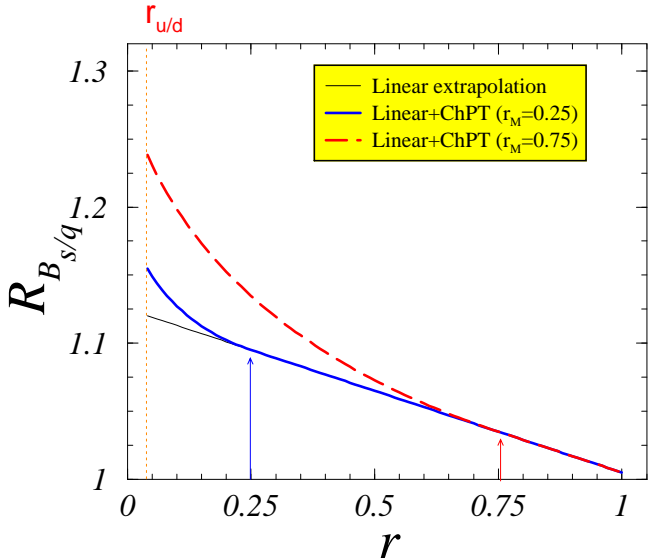}}
\caption[]{Illustration of the shift induced by the chiral logarithms in the extrapolation from the 
light quark masses accessed from the lattice, $r\in (0.5,1]$, to the physical point, $r_{u/d}=0.04$.  
The sensitivity of the resulting $f_{B_s}/f_{B_d}$ on the choice of the 
matching point $r_M$ [see eq.~(\ref{match})] is also shown. The phenomenon is illustrated 
by using $X_b=0.12(2)(3)$, and $g=0.5$.}
\label{fig4}\end{figure}

A common practice {\it ``to use a ruler"} and extrapolate the observed behaviour 
to the physical $r\to r_{u/d}$ is, however, dangerous because one is 
getting deeply into the region dominated by the spontaneous chiral symmetry 
breaking effects, so that the observed linear behaviour of the decay constants w.r.t. 
the change of the light quark mass may be significantly modified. 
That problem was first pointed out in ref.~\cite{sharpe}, and recently in~\cite{kronfeld-ryan}. 
It is therefore of paramount importance to do the computation with the light quark as light 
as possible, and check for the deviation from the form~(\ref{latt-slope}). In doing so, 
the finite volume effects should, of course, be kept under control. That is where the  
$\epsilon$-expansion should be applied in the way similar to what has been done recently in 
ref.~\cite{giusti}.~\footnote{It would be very interesting to 
actually verify the predictions of quenched ChPT, which trouble the 
lattice QCD community so much. Using the method of ref.~\cite{giusti}, one could check whether 
or not the (divergent) quenched chiral log term in the 
$f_K/f_\pi$ coincides with the QChPT prediction, 
$\displaystyle{{m_0^2\over 6(4\pi f)^2}{1+r\over 1-r}\log r}$, where 
$m_0\approx 0.65$~GeV, is the mass of the light quenched $\eta^\prime$ state.} Since such an unquenched study is not around the corner, we may 
rely on the chiral perturbation theory (ChPT) to guide the (chiral) extrapolations. 
By using the  Lagrangian in which the ChPT is combined with the static limit of HQET, 
the chiral logarithmic corrections to the SU(3) breaking ratio of the decay constants has been 
computed in~\cite{Grinstein} 
\bea\label{LOG1}
R_{B_{s/d}}^{\rm ChPT} \equiv {f_{B_s}\sqrt{m_{B_s}} \over f_{B_d}\sqrt{m_{B_d}}} =  1 \!\!\!&+&\!\!\! {1 + 3 {\hat g}^2\over 4 (4 \pi f)^2} \left(
3 m_\pi^2\log {m_\pi^2\over \Lambda_\chi} \right. \cr
&&\hspace*{-37.7mm}\left.- 2 m_K^2 \log {m_K^2\over \Lambda_\chi}- m_\eta^2\log {m_\eta^2\over \Lambda_\chi} \right) + {8 K(\Lambda_\chi
)\over f^2} (m_K^2 - m_\pi^2)  \,,
\eea
where $K$ stands for the unknown low energy constant.  
$R_{B_{s/d}}^{\rm ChPT}$ is independent of the chiral symmetry 
breaking scale $\Lambda_\chi\simeq 1$~GeV. This formula can be cast 
into the form ready for extrapolation by using the GMOR and Gell-Mann--Okubo 
formulae, 
\bea
m_\pi^2 = 2 B_0 m_s r\,,\quad m_K^2 = 2 B_0  m_s { r+1  \over 2}\,,\quad 
m_\eta^2 = 2 B_0 m_s { r+2 \over 3}\,.\nonumber 
\eea 
After noticing that $2B_0 m_s= 2 m_K^2 - m_\pi^2 = (684\ \mev)^2$, eq.~(\ref{LOG1}) 
can be written in terms of $r$, which we will then call $R_B^{\rm ChPT}(r)$.  
Notice that the last term is $\propto (1-r)$, much like in eq.~(\ref{latt-slope})  
in which the slope was obtained from the fit to the lattice data, and which we will refer to [eq.~(\ref{latt-slope}), that is] as 
$R_B^{\rm latt}(r)$. The main trouble actually lies in the fact that the chiral logarithms 
are multiplied by the {\it large} coefficient $(1+3 g^2)\approx 1.8$,~\footnote{
$\hat g$ is a coupling of the Goldstone boson to a doublet of the heavy--light mesons 
(e.g. $g_{B^\ast B\pi}$), whose determination will be discussed later on. } so that the application of the 
form $R_B^{\rm ChPT}(r)$ to extrapolate to $r_{u/d}$ results in a large shift, compared to 
what we obtain from the naive extrapolation, i.e. from the result of the application of $R_B^{\rm latt}(r_{u/d})$.
Since we do not directly observe the chiral logarithms on the lattice, the formula $R_B^{\rm ChPT}(r)$
can be used only below the region of the quark masses covered by our lattice simulations, namely for $r < r_M\simeq 0.5$. 
It can also be argued that $R_B^{\rm ChPT}(r)$ is not appropriate for the `pions' as heavy as $m_{``\pi"}= 0.48$~GeV
($\Leftrightarrow\ r_M=0.5$), and the form $R_B^{\rm latt}(r)$ can be extrapolated to lower $r_M$, before 
including the ChPT formula in the extrapolation. Clearly, the point $r_M$, at which the smooth matching 
can be made, 
\bea\label{match}
R_{B_{s/q}}(r)  &=& 
\vartheta(r -r_M ) R_{B_{s/q}}^{\rm latt}(r) \cr
&& + 
 \vartheta(r_M-r) \biggl[ R_{B_{s/q}}^{\rm ChPT}(r) - 
 \left(R_{B_{s/q}}^{\rm ChPT}(r_M)-R_{B_{s/q}}^{\rm latt}(r_M)\right)\biggr.\cr
 &&\biggl.  
 - \left( \left.{\partial R_{B_{s/q}}^{\rm ChPT}/\partial r}
\right|_{r=r_M}  + X_b\right) ( r - r_M) \biggr],
\eea
will result in different shifts. By taking $r_M=0.75$, a pure 
SU(3) breaking effect, $f_{B_s}/f_{B_d}-1$, gets enhanced by $100\%
$~\cite{kronfeld-ryan}. If, instead, we take $r_M=0.25$, the shift is still $+30 \%
$. This is illustrated in fig.~\ref{fig4}. An equivalent way to see that feature has been 
discussed in ref.~\cite{donoghue}, where the chiral loop integrals are computed by introducing the 
hard cut-off. The uncertainty on the value of the cut-off scale corresponds precisely 
to the uncertainty of the position in $r_M$ discussed above.
To get around that uncertainty,  before we are able to work with the light quarks sufficiently close to 
the chiral limit on the lattice, in ref.~\cite{sjsd} it was proposed to study the double ratio 
\bea\label{double}
&& R(r)= {R_{B_{s/q}}(r)\over f_{K_{sq}}/f_{\pi_{qq}}}\,.
\eea
There are two reasons why this is advantageous:
\begin{itemize}
\item[A.] The coefficients multiplying the chiral logarithms in 
\bea \label{rP}
R^{\rm ChPT}_{\pi_{u/d}} \equiv {f_{K}\over f_{\pi}} =   1 \!\!\!\!\!&+&\!\!\!\!\!  {1 \over 4 (4 \pi f)^2} \left[
5 m_\pi^2\log {m_\pi^2\over \Lambda_\chi} - 2  m_K^2\log {m_K^2 \over \Lambda_\chi}\right.\cr
&&\hspace*{-12.7mm}\left.- 3 m_\eta^2\log {m_\eta^2\over \Lambda_\chi} \right] + {8 L_5(\Lambda_\chi)\over f^2} (m_K^2 - m_\pi^2)  \,,
\eea
have the same sign and are almost of the same size as those appearing in 
eq.~(\ref{LOG1}).~\footnote{This statement obviously depends on the value of the 
coupling $\hat g$, which --in the $b$-quark sector-- I take to be $\hat g_b=0.52(7)(7)$, 
which will be discussed in the last part of this review.} 
Therefore, in the ratio, the logarithms will cancel to a large extent and the form
\bea
&&R(r)\ =\ A\ +\ B\cdot (1-r) \,
\eea
can be used in extrapolations, with a very small error due to the uncancelled logarithms. 
Using an equation similar to~(\ref{match}), but  for $R(r_{u/d})$, we notice a weak 
(negligible) sensitivity on the choice of $r_M$, because the large logarithms are cancelled. 
\item[B.] The physical result is obtained after multiplying $R(r_{u/d})$ 
by $f_K/f_\pi$, which is known from  experiment, namely $f_K/f_\pi = 1.22(1)$~\cite{PDG}. 
\end{itemize}
Concerning $f_K/f_\pi$, it is important to stress 
that, compared with the experimental value, the quenched lattice estimates 
obtained after extrapolating linearly,
\bea
&&R^{\rm latt}_{\pi}\ =\ 1\ +\ X_q \cdot (1-r)\,,
\eea
always lead to a small value. For example, 
from the SPQcdR data~\cite{spqr}, in the continuum limit, I get   
$X_q^{N_{\rm f}=0}=0.12(2)$, which then amounts to, $f_K/f_\pi -1=0.11(2)$, i.e.  
$50\%
$ smaller than the experimental value. The folkloric explanation for this ``spectacular 
failure" was/is the use of the quenched approximation. JLQCD collaboration 
recently showed that this is, however, not totally true. From their precision 
quenched and unquenched ($N_{\rm f}=2$)
computation, the results of which are reported in ref.~\cite{JLQCD-light}, 
I read off~\footnote{JLQCD also made a preliminary study with 
$N_{\rm f}=3$, indicating that $X_q^{N_{\rm f}=3}\approx X_q^{N_{\rm f}=2}$. }
\bea\label{slope-K}
&&X_q^{N_{\rm f}=0} = 0.125(7),\quad 
X_q^{N_{\rm f}=2} = 0.154\left(^{+16}_{-12}\right).
\eea 
The missing piece that would help getting to $(f_K/f_\pi)^{\rm exp.}$ 
is likely to be the one due to the chiral logarithms. This situation  
is precisely the opposite to the one that incited the ChPT practitioners 
to make the consistent NLO calculations, because the chiral logs alone give 
$f_K/f_\pi \simeq 1.14$, whereas at NLO the low energy constant $L_5$ [see eq.(\ref{rP})], 
is fixed so as to reproduce $(f_K/f_\pi)^{\rm exp.}$~\cite{GL}~.\footnote{As a side remark, one should 
stress that neither in ChPT nor in QCD sum rules the $f_K/f_\pi$ is obtained directly, but rather 
after fixing the extra-parameters, i.e. not only changing the quark mass $m_{u,d}\leftrightarrow m_s$   
(for the recent QCDSR analysis of $f_K/f_\pi$, please see ref.~\cite{mannel}). 
This is why a thorough lattice study, allowing a deeper understanding of the SU(3) breaking mechanism, is 
badly needed.} On the lattice, instead, we see the explicit linear quark mass 
dependence but there is still no clear evidence for the presence of 
the chiral logs (most probably because we are not working with sufficiently 
light quarks). 
To get the lattice estimate of $f_K/f_\pi$, we then face the same 
problem as before: chiral logs are large and the results of extrapolation 
by using an expression similar to eq.~(\ref{match}) would be strongly 
dependent on the choice of the matching point $r_M$.

The double ratio~(\ref{double}) avoids all those headaches, and can be computed on the lattice 
directly. By combining $X_q= 0.154\left(^{+16}_{-12}\right)$, with $X_b =0.15(4)$, I get 
\bea\label{FB}
{f_{B_s}\over f_{B_d}}&=& {f_K^{\rm exp}\over f_\pi^{\rm exp}}\left[  {1\ +\ X_b\cdot r_{u/d}
\over 1\ +\ X_q\cdot r_{u/d}} +  \sqrt{m_{B_d}\over m_{B_s}} \times {\rm ``chi-logs"}_{\hat g}\right]\cr
&& \hfill \cr
&=& 1.21\pm 0.05\pm 0.01\,,
\eea
where the last error is due to the variation of the coupling $\hat g= 0.52(7)(7)$. 
A similar proposal, to consider $\sqrt{m_B}f_B/f_\pi$, has been made in ref.~\cite{milc}.  
However, one has to assume that the terms of ${\cal O}(1/m_b)$, which are known to be large, 
do not induce any extra chiral log dependence. Large ${\cal O}(1/m_b)$
correction cancel in the $f_{B_s}/f_B$ ratio.

A completely analogous procedure to the one sketched above applies to the charm sector. 
I will use $X_c = 0.11(4)$, which is obtained from $X_c^{N_{\rm f}=0}=0.09(1)(1)$~\cite{nazario}, and 
the observation made in refs.~\cite{CPPACS2,milc}, namely  
$(f_{D_s}/f_{D})^{N_{\rm f}=2}/(f_{D_s}/f_{D})^{N_{\rm f}=0}=1.02(3)$. 
Together with the experimentally measured $\hat g_c=0.61(6)$, and 
from the formula analogous to eq.~(\ref{FB}), I arrive at ${f_{D_s}/f_D} = 1.22(4)$.  

In summary,\\
\psshadowbox[fillcolor=yellow,linewidth=.0cm]{{\parbox[c]{8.1cm}{
\bea
&&{f_{D_s}\over f_D} = 1.22 \pm 0.04\,,\cr
&& \hfill \cr
&&{f_{B_s}\over f_{B}} = 1.21 \pm 0.05\pm 0.01\,,
\eea}}}\\

\noindent
or, within the error bars, $f_{B_s}/f_{B} \approx f_{D_s}/f_{D} \approx f_K/f_\pi$.

\section{$\bbar$ mixing and $\xi$ }

The bag parameter that parameterizes the $\bbar$ mixing amplitude is defined by
\bea
 \label{Bparam}
\langle \bar B^0_q \vert \bar b^i \gamma_\mu (1- \gamma_{5} )  q^i \,
 \bar b^j  \gamma_\mu (1- \gamma_{5} ) q^j \vert   B^0_q \rangle  
= {8\over 3} \, m_{B_q}^2  f_{B_q}^2
\hat B_{B_q} \ ,  
\eea
i.e. normalized to the vacuum saturation approximation ($B_{VSA}=1$). 
This amplitude is very difficult to calculate by QCD sum rules because: (i) the calculation of 
the NLO corrections to the perturbative part of the spectral function, as well as to the 
Wilson coefficients multiplying the condensate contributions, is quite involved; 
(ii) the stability of the sum rule under the variation of the threshold and the Borel 
parameters is very hard to achieve. Some progress concerning 
the first part of the problem was made recently in ref.~\cite{petrov}. So far, the 
lattice QCD studies of the matrix element~(\ref{Bparam}) are made by 
using the Wilson quark action. The unpleasant feature of the Wilson quarks is the lack of explicit chirality,
 which induces complications in renormalization, namely it allows the operator 
in eq.~(\ref{Bparam}) to mix with  other parity-even, $\Delta B=2$, dimension-six operators. 
The extra mixing is the lattice artefact and should be subtracted, preferably non-perturbatively. 
To a discussion provided in the Yellow Book from the previous CKM-workshop~\cite{yellowbook}, I would 
like to add the recent proposal made in ref.~\cite{withJuan}, to combine the overlap light 
quark action (which preserves the chirality) with HQET, which enormously simplifies the renormalization
procedure. 
\begin{figure}[tbh]
\vspace*{-2mm}\hspace{+3.5mm}\centerline{{\epsfxsize=0.22\textwidth\epsffile{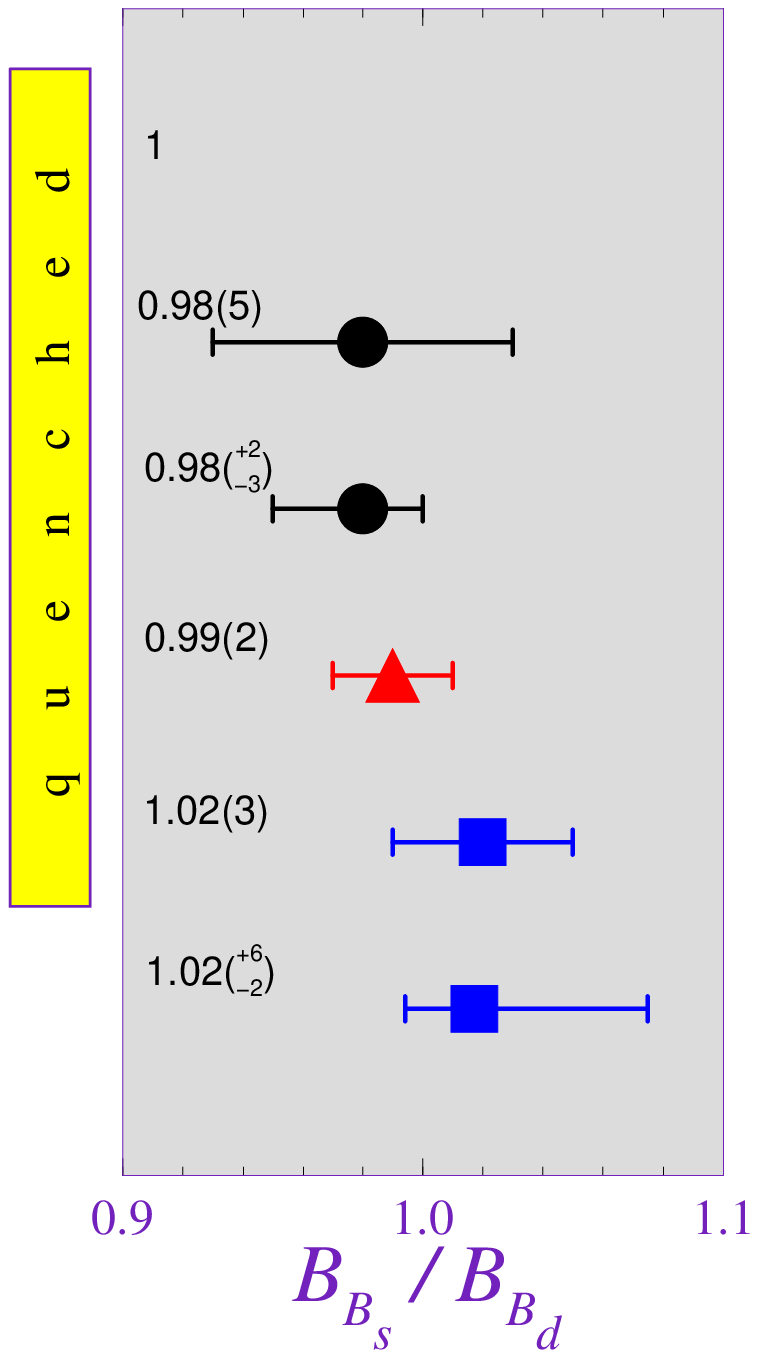}}{\epsfxsize=0.35\textwidth\epsffile{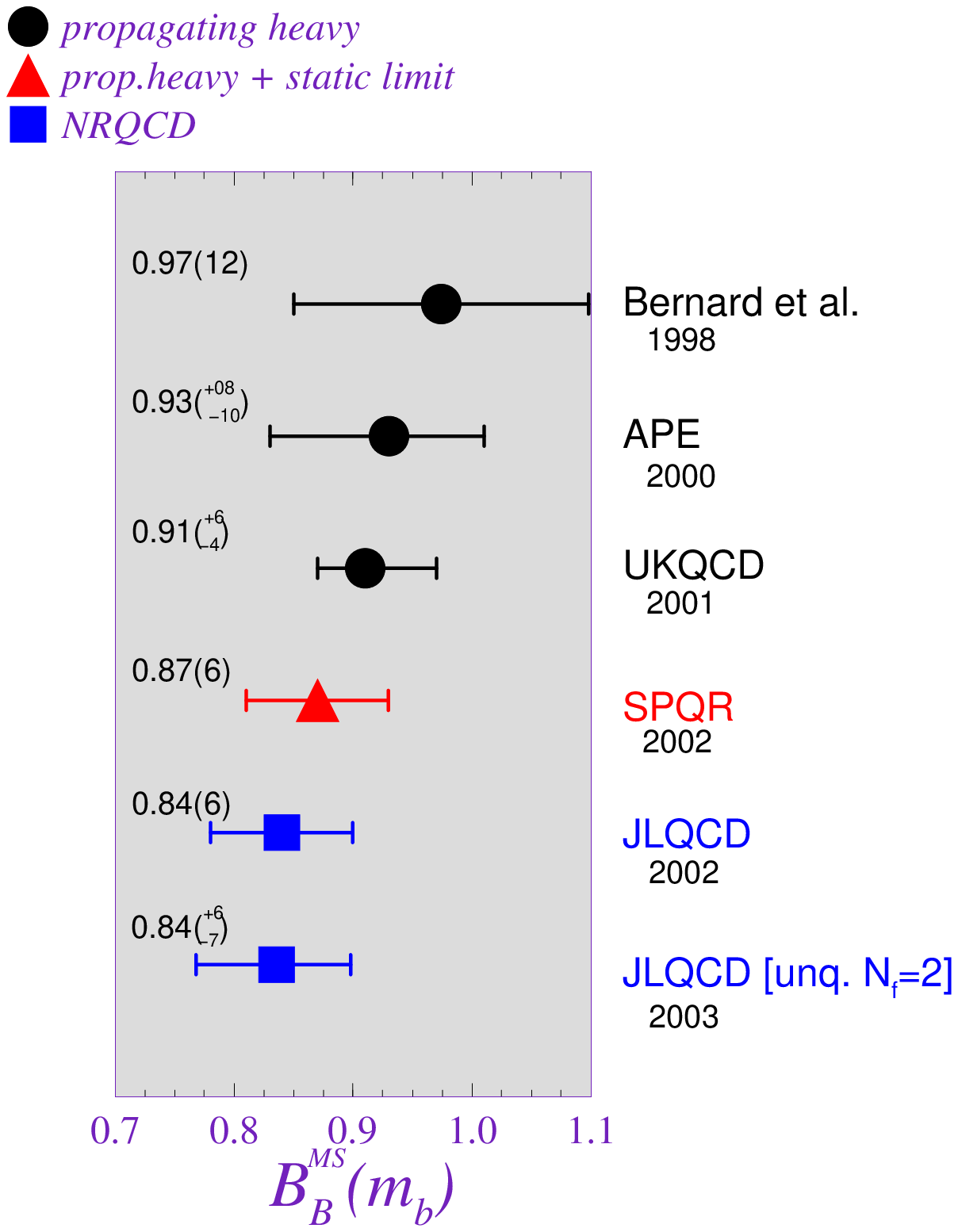}}}
\caption[]{SU(3) breaking ratio of the bag parameters and $B_{B_d}^\msbar (m_b)$, as obtained from 
the lattice studies by extrapolating from the accessible propagating heavy quarks, by constraining the extrapolation 
with the static HQET result, and by using the NRQCD treatment of the heavy quark. The plotted results are taken from 
refs.~\cite{SONI,APE,LL,db2APE,JLQCD2,JLQCD3}.}
\label{fig5}\end{figure}

In fig.~\ref{fig5}, we show the results for the SU(3) breaking ratio of the bag parameter, 
as well as the absolute value of $B_B(m_b)$ in the $\msbar$(NDR) scheme. Let us briefly discuss 
some important points. In ref.~\cite{db2APE}, the static HQET result has been combined with the 
ones obtained with the propagating heavy quarks. That was an important progress because, 
instead of extrapolating, one could interpolate to get the desired result. In doing so 
the matching of the QCD results with the static (HQET) ones has been made at NLO in perturbation 
theory, which is important for two reasons:(1) the anomalous dimensions and mixing patterns are different in 
the full and effective theories (QCD and HQET), and to use the heavy quark scaling laws the matching 
to HQET is required; (2) to cancel the scale and scheme dependence against the perturbatively computed 
Wilson coefficient~\cite{buras} [i.e. $C_B(\mu)$ in eq.~(\ref{UTA})], one has to provide $B_{B_{s,d}}(\mu)$, 
computed in the same $\msbar$(NDR) renormalization scheme. All schemes are equal at the leading order, 
and to specify the renormalization scheme, one must go beyond, i.e. to NLO. All that is done consistently in 
ref.~\cite{db2APE}. However, those results are quenched, they are obtained at a single value of the lattice spacing, 
and the matrix elements in HQET were renormalized  only perturbatively. 
Important step in taming the quenching effects has been made in refs.~\cite{JLQCD2,JLQCD3}, 
where the heavy quark is treated non-relativistically. From the comparison of the results obtained in simulations with 
$N_{\rm f}=0$ and $N_{\rm f}=2$, they see no difference between 
$B_{B_{d/s}}^{N_{\rm f}=0}$ and $B_{B_{d/s}}^{N_{\rm f}=2}$.

We should reiterate that the light quark in $B_{B_s}(\mu)$ is accessed directly, 
whereas $B_{B_d}(\mu)$ is reached through an extrapolation. ChPT suggests no deviation 
of the SU(3) breaking ratio from the linear form. 
From ref.~\cite{Grinstein} we know that the chiral log term reads
\bea\label{Bfac}
 {\hat B_{B_s} \over \hat B_{B_d}} =
1 + {\overbrace{1 -3 {\hat g}^2}^{\displaystyle{tiny!}} \over 2 (4 \pi f)^2} \left[   
m_\pi^2\log m_\pi^2 -  m_\eta^2\log m_\eta^2 \right] \;,
\eea 
which, contrary to the case of the decay constants, is very small; no significant shift in the chiral 
limit is therefore to be expected. That gives us confidence that the chiral extrapolations do not induce 
important systematic uncertainties, although that issue will be resolved {\it iff} we are able to actually do the 
lattice computation close to the chiral limit. By taking the simple average of the results given in fig.~\ref{fig5}, 
we obtain $B_B^\msbar (m_b)=0.89(3)$, to which we add $10\%
$ of uncertainty due to possible remaining: quenching, chiral extrapolation, continuum extrapolation effects. 
To convert from $\msbar$ to RGI form, with $\Lambda_{\rm QCD}^{(5)}=213(32)$~MeV~\cite{bethke}, 
the conversion factor in $\hat B_{B_{d/s}} = c(m_b) B_{B_{d/s}}(m_b)$, is 
$c^\msbar(4.8\ \gev)=1.54(1)$. The average SU(3) breaking ratio of the bag parameters gives 
$1.00(1)$, the error of which we double for the same reasons mentioned above. Therefore we have

\psshadowbox[fillcolor=yellow,linewidth=.0cm]{{\parbox[c]{8.1cm}{
\bea
\hat B_B = 1.37(14),\;B_{B_s}/B_B = 1.00(2) ,\;\xi = 1.21(6)\ .
\eea}}}\\

Since there is still much room for the improvement of the bag parameter results, I should also 
mention the recent QCDSR result, $B_B^\msbar (m_b)\sim 0.95$~\cite{petrov}, as well as the one of the 
chiral quark model,  $B_{B_s}/B_{B_d} = 0.93(5)$~\cite{eeg}.~\footnote{From the plethora of predictions given in 
ref.~\cite{eeg}, I chose to quote the one for $B_{B_s}/B_{B_d}$,  which is 
very stable under large variations of the model parameters.}

\section{$\hat g$ coupling}

The coupling of the lowest lying doublet of heavy--light mesons ($j_\ell^P =1/2^-$) to a charged soft pion, 
$g_{H^\ast H \pi}$, is defined as
\bea
&&\langle H(p^\prime )\pi(q)\vert H^\ast (p)\rangle = g_{H^\ast H \pi} \ \left( q\cdot \varepsilon(p)\right)\;.
\eea 
This coupling appears to be the essential parameter in the chiral Lagrangian of 
the heavy--light meson systems, ${\cal L}_{\rm int}=\hat g {\rm Tr}[\bar H H \gamma_\mu \gamma_5 {\bf A}^\mu]$, 
with $\hat g \propto g_{H^\ast H \pi}$. The constant $g_{H^\ast H\pi}$ is important in describing the 
$D\to \pi \ell \nu$ and $B\to \pi \ell \nu$ decay form factors, since the residuum of the dominant form 
factor at the nearest pole (i.e. $m_{D^\ast}^2$ and $m_{B^\ast}^2$, respectively) is given by 
$f_{H^\ast}m_{H^\ast}g_{H^\ast H \pi}/2$. 
The value of this coupling has  recently been measured by CLEO in the charm sector, 
$g_{D^\ast D\pi}= 17.9\pm 0.3\pm 1.9$~\cite{CLEO}, which we then convert into the value that is expected 
to scale with the heavy quark mass as a constant (up to $1/m_Q$ corrections and higher), namely
\bea
&&g_{D^\ast D \pi} = { 2 \sqrt{m_D m_{D^\ast}}\over f_\pi } \hat g_c, \quad \hat
g_c^{\rm exp.}= 0.59(7)\,,
\eea 
where the index ``$c$" indicates that the heavy quark is charm; $\hat g_b$, on the other hand, cannot be 
measured directly because there is no available phase space for the pion emission (or absorption). 
A summary of the predictions of this quantity as obtained by using various approaches can be found in 
ref.~\cite{singer}. Here I will focus on the recent development.

The experimental finding by CLEO was surprising in that the value for $g_{D^\ast D \pi}$ 
was much larger than the one predicted by the light cone QCD sum rules (LCSR)~\cite{khodjamirianG}. 
To cure that discrepancy, ref.~\cite{quark} proposed to include the  first radial excitations in 
the hadronic sum, i.e. below the continuum threshold in the double dispersion relation. 
As a result, the LCSR value for $g_{D^\ast D\pi}$ becomes much larger and more stable against the 
variation of the sum rule parameters. In addition, the sum rule for the $D^{(\ast)}$-meson decay 
constant remains unchanged, whereas the $D\to \pi\ell \nu$ form factor gets corrected in such a way that 
the result of pole dominance becomes closer to the experimental value for this form factor. 

\begin{figure}[tbh]
\vspace*{-5mm}\centerline{\epsfxsize=0.4\textwidth\epsffile{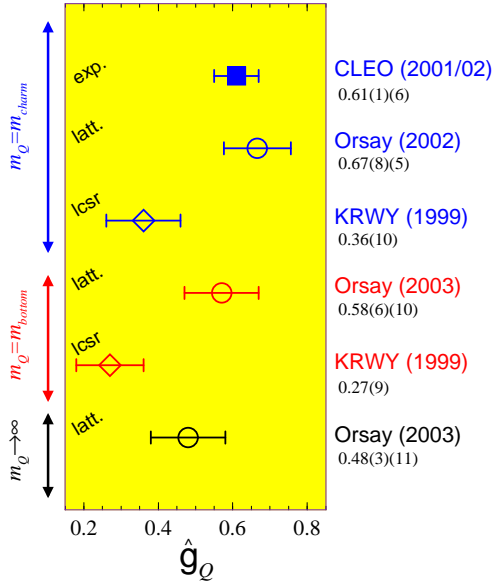}}
\caption[]{Present situation of the estimates of the $\hat g$-coupling between 
the lowest doublet of heavy--light mesons and the pion. Plotted are 
the results obtained by using the light cone QCD sum rules (LCSR)~\cite{khodjamirianG}, (quenched) 
lattice QCD~\cite{orsay,orsay2} and the experimental value~\cite{CLEO}. }
\label{fig6}\end{figure}

The first lattice computation of this coupling 
was made last year~\cite{orsay}. The findings of that reference can be summarized as 
follows:
\begin{itemize}
\item[1.] From the simulations in the quenched approximation, it appears that the finite lattice spacing 
and finite volume effects are small;
\item[2.] The coupling is computed from the transition form factor between the light quarks in $H$ and $H^\ast$, via the
axial current, with the heavy quark being only a spectator. The coupling $\hat g_Q$ is almost insensitive 
to the value of the heavy quark mass for the heavy masses around the charm quark (see fig.5 of ref.~\cite{orsay});
\item[3.] The resulting value is $\hat g_c= 0.67(8)(^{+4}_{-6})$, i.e. it is large and in 
good agreement with $\hat g_c^{\rm exp.}$.
\end{itemize}
Concerning point 2, the observation that the slope  in
\bea
&&\hat g_Q =\hat g_\infty\ \left( 1  + \gamma/m_Q \right)\; 
\eea
is $\gamma^{\rm latt}\approx 0$, disagrees with the LCSR prediction, $\gamma^{\rm lcsr}\approx 
0.6$~GeV. It is important to say that LCSR came to that conclusion by computing the couplings $\hat g_b$ and $\hat g_c$, 
while on the lattice only the masses close to the charm quark are studied. 
Since $\hat g_b$ is highly important 
[see e.g. eq.~(\ref{LOG1})], it is necessary to have a better control of the slope $\gamma$. 
To get around that problem, the Orsay group very recently followed the strategy of ref.~\cite{ukdipierro} 
and computed $\hat g$ on the lattice in the static limit of HQET.~\footnote{
In ref.~\cite{orsay2}, the recipe to better the statistical quality of the signal by fattening the link variables 
in the Wilson line has been implemented.}  Orsay group obtains $\hat g_\infty = 0.48(3)(10)$, where the bulk 
of systematic uncertainty stems from various ways of smearing the source operators (the ones that 
produce $H^{(\ast)}$)~\cite{orsay2}. Thus the result for the slope is very close to the one predicted 
by LCSR, $\gamma^{\rm lcsr}$, although the absolute value obtained on the lattice and from LCSR do disagree. 
With the static result in hand, one can interpolate to the $B$-meson mass, which results in 
$\hat g_b=0.58(6)(10)$.~\footnote{ 
In estimating the chiral logs effects in the previous section, I used 
$\hat g_b= 0.52(7)(7)$, totally consistent with the new result by the Orsay group. }
The present situation of the estimates based on LCSR and quenched lattice simulations is shown in fig.~\ref{fig6}. 
Besides the obvious necessity to go beyond quenching and to better control the chiral extrapolations, 
it would be nice if other lattice groups produced results for this coupling, even in the quenched approximation.

\section*{Acknowledgements}
I would like to thank all my collaborators, the authors of refs.~\cite{rolf1,nazario} for correspondance, 
the organizers for inviting me to this useful workshop and the E.U.'s financial support 
under contract HPRN-CT-2000-00145 ``Hadron Phenomenology from Lattice QCD".

\end{document}